\documentclass[preprint,article,accept,moreauthors,pdftex]{Definitions/mdpi_pre}

\firstpage{1} 
\pubvolume{xx}
\issuenum{1}
\articlenumber{0}
\pubyear{2020}
\copyrightyear{2020}
\externaleditor{Academic Editor: name}
\history{Received: date; Accepted: date; Published: date}

\usepackage{natbib}

\usepackage{subcaption}

\usepackage{wrapfig}

\Title{Analytical approach to solve the problem of aircraft passenger boarding during the coronavirus pandemic}

\Author{Michael Schultz $^{1,\dagger}$ and Majid Soolaki $^{2,\dagger}$}
\AuthorNames{Michael Schultz and Majid Soolaki}

\address{%
$^{1}$ \quad Institute of Logistics and Aviation, Dresden University of Technology\\
$^{2}$ \quad School of Mechanical and Materials Engineering, University College Dublin}

\firstnote{These authors contributed equally to this work.} 

\abstract{The handling processes at the airport will be significantly changed as a consequence of the corona epidemic. We expect pandemic requirements will be establish as permanent as the inherent requirements for safety and security in aviation. During the aircraft boarding, passengers are near each other, which requires both an effective rule to guarantee physical distances and an efficient procedure to obtain appropriate boarding times. We design an optimal group boarding method using a stochastic cellular automata model for passenger movements, which is extended by a virus transmission approach. Furthermore, a new mathematical model is developed to determine an appropriate seat layout for groups. The proposed seating layout is based on the idea that group members are allowed to have close contact and that groups should have a distance among each other. The sum of individual transmission rates is taken as the objective function to derive scenarios with a low level transmission risk. After the determination of an appropriate seat layout, the cellular automata is used to derive and evaluate a corresponding boarding sequence aiming at both short boarding times and low risk of virus transmission. We find that the consideration of groups in a pandemic scenario will significantly contribute to a faster boarding (reduction of time by about 60\%) and less transmission risk (reduced by 85\%), which reaches the level of boarding times in pre-pandemic scenarios.}


\begin{document}


\section{Introduction}
\label{sec:intro}
Air transportation offers international mobility in a globally well-connected network. In view of the current pandemic situation caused by the new coronavirus SARS-CoV2, air transportation is part of the transmission chains and is sustainably affected \cite{ICAO_2020}. Along the passenger journey, the cabin operations demand to share a constricted environment with other passengers during boarding, flight, and deboarding. Thus, these operations hold the potential for virus transmissions between passengers and require an appropriate seat allocation and strategies to reduce the transmission risk significantly. The constraints from pandemic situations lead to changes in passenger handling. Currently, airlines aim to protect passengers and crews from Covid-19 and see face covering as mandatory for passengers onboard. A physical distance between passengers during aircraft boarding and deboarding is also part of airlines risk mitigation strategies. With our contribution we provide a customer-oriented solution for both airlines and passengers, which enables a situative approach to establish appropriate seat layout and aircraft entry sequences considering minimum interactions between groups of passengers.

\subsection{Review of research on virus transmission in aircraft}  
The SARS outbreak in 2002 emphasizes the important role of air transportation in pandemic situations \cite{SARS_2003}. The climate control system of aircraft seems to reduce the spreading of airborne pathogens by frequently recirculating the cabin air through high efficiency particulate air (HEPA) filters \cite{mangili_transmission_2005,IATA_MedEvidence_2020}. These filters are designed to filter at least 99.95\% of aerosols and are capable of removing viruses and bacteria attached to droplets. But the transmission of infectious diseases is likely to be more frequent than reported for several reasons, such as a much shorter flight times than incubation periods. When considering the passenger path to and in the aircraft cabin, upstream and downstream processes can also lead to infections (e.g. baggage handling, security checks). To minimize physical interactions, current handling approaches aim at a contactless passenger journey through the airport terminal. In a post-pandemic scenario this contactless journey could include biometric scans or the use of personal mobile devices for services or inflight entertainment.  Studies on reported in-flight transmissions emphasize that proximity to the index case increases the risk of transmission \cite{SARS_2003,2RowRule_2016}. The simulation of transmissions during flight, based on actual passenger behaviors in single-aisle aircraft, indicate a low probability of direct transmission to passengers not seated in close proximity to an infectious passenger \cite{Hertzberg3623}. An investigation of a long-haul flight indicates a low risk of pandemic influenza transmission close to infected passengers with symptoms \cite{baker_transmission_2010}. The calculation of the spatial and temporal distributions of droplets in an aircraft cabin showed a reduced influenza transmission risk, if respirator masks are used by the passengers \cite{gupta_risk_2012}. The documentation of a symptomatic SARS-CoV2 index case flying a 15 hour trip in economy class shows that all 25 passengers being seated within a range of 2~m from the index case were tested negative for SARS-CoV2 \cite{SARSCoV2_20202}. Two other case studies reports 11 transmissions \cite{China_transmission_2020} and one potential infection during a flight \cite{France_transmission_2020}. A brief introduction about the understanding of~SARS-CoV2 in the context of passenger boarding is given at \cite{schultz_evaluation_2020}.

\subsection{Review of aircraft boarding approaches}
Comprehensive overviews are provided for aircraft ground operations, passenger boarding, and corresponding economic impact \cite{schmidt_review_2017,jaehn_airplane_2015,_Schultz2018c,nyquist_study_2008,mirza_economic_2008,cook_european_2015,delcea_agent-based_2018,_Schultz2008a}. A common goal of simulation-based approaches for passenger boarding is to minimize boarding time. Thus, the efficiency of different boarding strategies was focus of the research activities \cite{marelli_role_1998,van_landeghem_reducing_2002,ferrari_robustness_2005,van_den_briel_america_2005,bachmat_bounds_2008,_Schultz2008b,bachmat_optimal_2013,steffen_optimal_2008}. These models are based on cellular automaton or analytical approaches, but also other models were developed: mixed integer linear programs \cite{bazargan_linear_2007,majid_soolaki_2012}, statistical mechanics \cite{steffen_statistical_2008}, power law rule \cite{frette_time_2012,bernstein_comment_2012}, cellular discrete-event system specification \cite{jafer_comparative_2017}, stochastic approach covering individual passenger behavior and aircraft/airline operational constraints \cite{_Schultz2008b,_Schultz2018c}.

The quantity and quality of hand luggage determine the duration of boarding significantly. Thus, research was conducted with a particular focus on the physique of passengers (maximum speed), the quantity of hand luggage, and individually preferred distance \cite{tang_aircraft_2012}, seats assigned to passengers with regards to hand luggage \cite{qiang_reducing_2014,milne_optimization_2016,steffen_optimal_2008,milne_new_2014}. Furthermore, the fact that passengers travel in groups has an impact on the boarding efficiency \cite{zeineddine_dynamically_2017,_Schultz2018c}. Other research aims at the evaluation of pre-boarding areas \cite{steiner_speeding_2009,wallace_flying_2013}, consideration of passenger expectations \cite{wittmann_customer-oriented_2019}, use of apron busses \cite{milne_new_2019}, and real-time seat allocation \cite{_Schultz2018b,yazdani_real-time_2019}. The impact of different aircraft cabin layouts on passenger boarding were focused on the following studies: aircraft interior design (seat pitch and passengers per row) \cite{bachmat_analysis_2009}, aircraft seating layouts and alternative designs single and twin-aisle configuration \cite{_schultz2013b,chung_simulation_2012}, impact of aircraft cabin modifications \cite{fuchte_enhancement_2014}, novel aircraft configurations and seating concepts \cite{schmidt_novel_2015, schmidt_boarding_2017}, and dynamic change of the cabin infrastructure \cite{_schultz2017a}. Only few experimental tests have been conducted to provide data for the calibration of input parameters and validation of simulation results: using a mock Boeing 757 fuselage \cite{steffen_experimental_2012}, time to store hand luggage items in the overhead compartments \cite{kierzkowski_human_2017}, small-scale laboratory tests \cite{gwynne_small-scale_2018}, evaluation of passenger perceptions during boarding/deboarding \cite{miura_passenger_2017}, operational data and passenger data from field trial measurements \cite{_Schultz2017i,_Schultz2018i}, field trials for real-time seat allocation in connected aircraft cabin \cite{_Schultz2018b}, and using a B737-800 mock-up (1/3 size) to explore the factors effecting the time of luggage storage \cite{ren_new_2020}. 

There are two new research contributions available, which set a focus on behaviors during pandemic situations and their impact on the aircraft boarding procedures. The first research addresses the quantity and quality of passenger interactions \cite{cotfas_evaluating_2020} and the second research additionally develops and implements a transmission model to provide a more detailed evaluation \cite{schultz_evaluation_2020}. With a focus on airport operations, the impact of physical distances on the performance of security control lanes was analyzed to provide a reliable basis for appropriate layout adaptations \cite{kierzkowski_simulation_2020}.

\subsection{Focus and structure of the document}

We provide in this contribution an approach for aircraft boarding considering pandemic scenarios. These scenarios are mainly driven by the requirement of physical distance between passengers to ensure a minimal virus transmission risk during the boarding, flight, and deboarding. We consider passenger groups as an important factor to derive an appropriate seat layout and boarding sequence. The main idea behind the group approach is that members of one group are allowed to be close to each other, as they are already in close contact with each other before boarding, while different groups should be separated as far apart as necessary. Deboarding is not explicitly considered in our contribution. The paper is structured as follows. After the introduction (Sec.~\ref{sec:intro}), we present a stochastic cellular automata approach, which is used for modeling the passenger movements in the aircraft cabin (Sec.~\ref{sec:pax_boarding}). A transmission model is implemented to evaluate the virus transmission risk during passenger movements and applied to evaluate standard boarding procedures. In Sec.~\ref{sec:optimization}, we introduce a problem description and optimization strategies considering passenger groups. The results of the optimization model are presented in Sec.~\ref{sec:application_optimization}, where we use a genetic algorithm for solving the complex problem. The achieved seat layouts are used as input for the passenger movement model to derive an appropriate boarding sequence with a minimized transmission risk during boarding. Finally, our contribution ends with conclusion and outlook (Sec.~\ref{sec:discussion}).

\section{Passenger boarding model using operational, individual, pandemic constraints}
\label{sec:pax_boarding}

The initial model for movements of pedestrians was developed to provide a stochastic approach covering short (e.g. avoid collisions, group behavior \cite{_Schultz2013c}) and long-range interactions (e.g. tactical wayfinding) of human beings \cite{_Schultz2010}. This cellular automata model is based on an individual transition matrix, which contains the transition probabilities to move to adjacent positions around the current position of the passenger \cite{_Schultz2013d}.

\subsection{Operational constraints and rules of movement}
\begin{figure}[htb!]
    \centering
    \includegraphics[width=.7\textwidth]{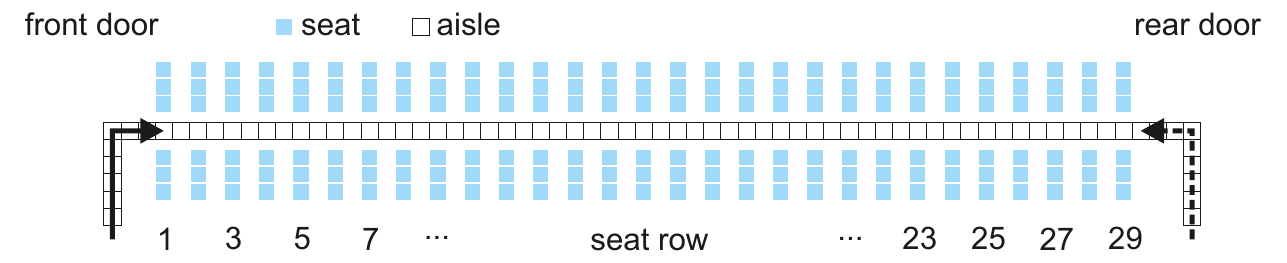}
    \caption{Grid-based model - Airbus A320 with 29 seat rows and 6 seats per row (reference layout).}
    \label{fig:a320_grid}
\end{figure}

To reflect operational conditions of aircraft and airlines (e.g. seat load factor, conformance to the boarding procedure) as well as the non-deterministic nature of the underlying passenger processes (e.g. hand luggage storage) the stochastic model was developed \cite{_Schultz2008b,_schultz2013b} and calibrated \cite{_Schultz2017i,_Schultz2018i}. The model will be used for the passenger movements during the aircraft boarding. The passenger boarding is modeled with a cellular automata approach based on a regular grid (Fig.~\ref{fig:a320_grid}). This regular grid consists of equal cells with a size of 0.4 x 0.4 m, whereas a cell can either be empty or contain exactly one passenger.

The boarding progress consists of a simple set of rules for the passenger movement: a) enter the aircraft at the assigned door (based on the current boarding scenario), b) move forward from cell to cell along the aisle until reaching the assigned seat row, and c) store the luggage (aisle is blocked for other passengers) and take the seat. The storage time for the hand luggage depends on the individual number of hand luggage items. The seating process depends on the constellation of already used seats in the corresponding row. A scenario is defined by the seat layout, the number of passengers to board, the arrival frequency of the passengers at the aircraft, the number of available doors, the boarding strategy and the conformance of passengers in following the current strategy. Further details regarding the model and the simulation environment are available at \cite{_Schultz2018c}. 

In the simulation environment, the boarding process is implemented as follows. Depending on the seat load, a specific number of randomly chosen seats are used for boarding. For each seat, an agent (passenger) is created. The agent contains individual parameters, such as the number of hand luggage items, maximum walking speed in the aisle (set for all agents to 0.8 m/s \cite{_Schultz2018i,_Schultz2018b}), seat location, time to store the hand luggage and arrival time at the aircraft door (both stochastically distributed). The agents are sequenced with regard to the active boarding strategy. From this sequence, a given percentage of agents are taken out of the sequence (non-conformant behavior) and inserted into a position, which contradicts the current strategy (e.g. inserted into a different boarding block). 

A waiting queue at the aircraft door is implemented and each agent enters this queue at the arrival time. In each simulation step, the first agent of the queue enters the aircraft by moving to the entry cell of the aisle grid (aircraft door), if this cell is free. Then, all agents located in the aisle move forward to the next cell, if possible (free cell and not arrived at the seat row), using a shuffled sequential update procedure (emulate parallel update behavior \cite{_Schultz2010,_Schultz2013d}). If the agent arrives at the assigned seat row, the corresponding cell at the aisle is blocked until the hand luggage is stored. Depending on the seat row condition (e.g. blocked aisle or middle seat or both), additional time is added to perform the seating process (seat shuffle). When the seating process is finished the aisle cell is set free. Each boarding scenario is simulated 125,000 times, to achieve statistically relevant results defined by the average boarding time (starts when the first passenger arrives the aircraft and finished when the last passenger is seated) and the standard deviation of boarding times.

\begin{figure}[htb!]
    \centering
    \includegraphics[width=.95\textwidth]{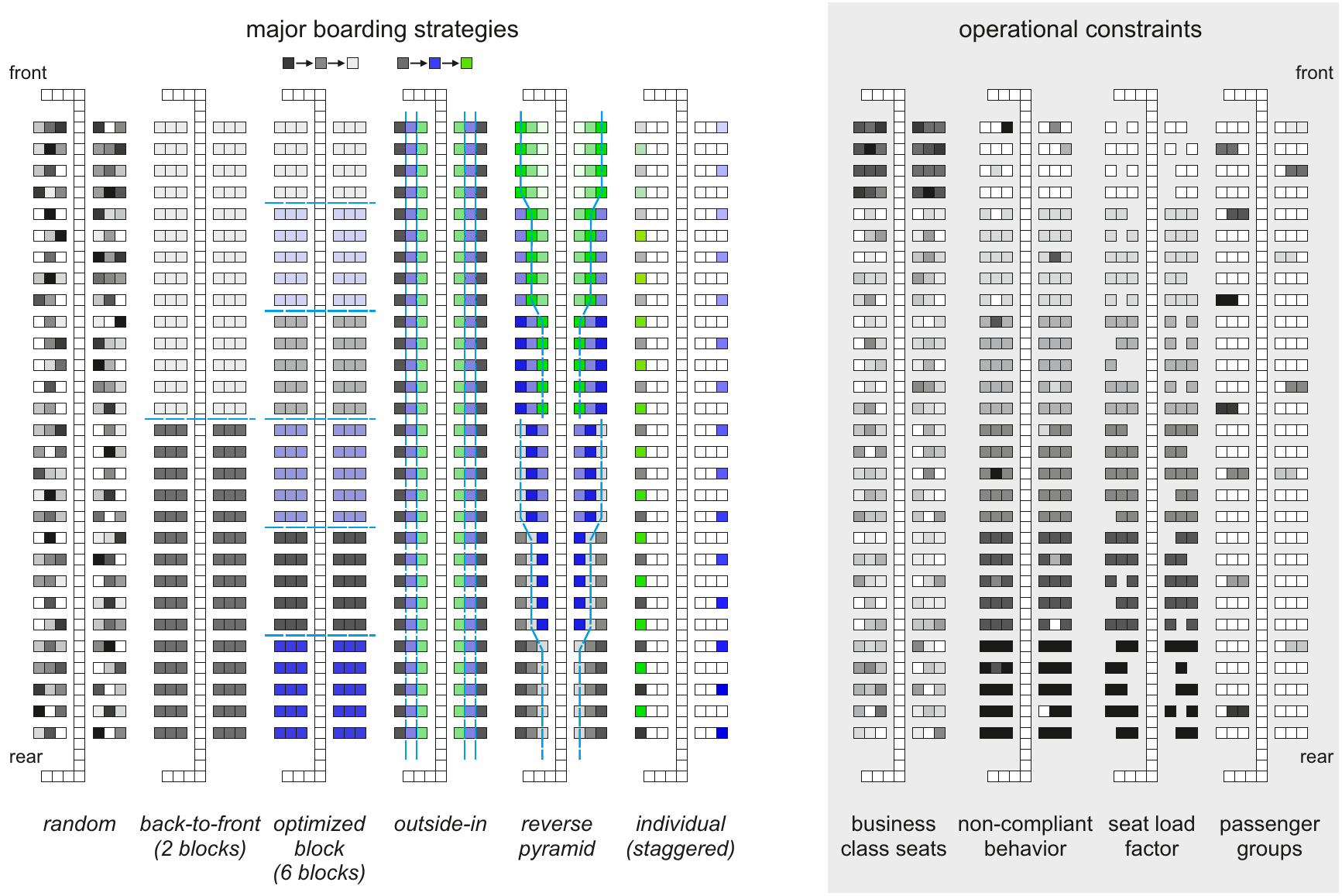}
    \caption{Overview of different boarding strategies (darker seats are boarded first; black - blue - green) and implementation of operational constraints in the cellular automata model.}
    \label{fig:boarding_strategies}
\end{figure}

Boarding strategies are derived from three major approaches: boarding per rows (aggregated to blocks), boarding per seat (window, middle, aisle), and sequences of specific seats. Fig.~\ref{fig:boarding_strategies} (left) depicts how the boarding strategies and operational constraints are implemented in the boarding model. The seats are color-coded to emphasize the order of aircraft seats in the boarding sequence. Six different boarding strategies are generally considered: random, back-to-front (based on 2 blocks), optimized block (based on 6 blocks), outside-in (window seats first, aisle seats last), reverse pyramid (back-to-front plus outside-in with 6 blocks), and individual seating. 

Thus, boarding strategies range from random boarding without a specific order to individual boarding, which is a specific solution of the optimized block (alternating seats) and the outside-in strategy (each block contains only one seat). Fig. \ref{fig:boarding_strategies} (right) illustrates how the operational constraints of 1$^\text{st}$ class seats, passenger conformance, seat load factor, and the existence of passenger groups are covered by the boarding model.

\subsection{Transmission model}

The fundamental cellular automata developed for the stochastic passenger movements is extended by an approach to evaluate the risk of a virus transmission during the boarding process. We are not considering face masks in our approach. The transmission risk can be defined by two major input factors: distance to the index case and reduction of contact time. A straight forward approach is to count both the individual interactions (passengers located in adjacent cells) and the duration of these contacts in aisle and during the seating process. However, counting the individual contacts will only provide a first indication about potential ways of infections. 

We derived a more comprehensive approach, which is based on the transmission model~\citep{smieszek_epidemicsmodel_2009} defining the spread of SARS-CoV2 as a function of different public distancing measures~\citep{nagel_mobilityberlin_2020}. The probability of a person $n$ to become infected in a time step $t$ is described in Equation~\eqref{eq:mechanistic_model}.

\begin{equation}
    P_{ n,t } = 1 - exp \left( -\theta \sum \ \text{SR}_{ m,t } \quad i_{ nm,t } \quad t_{ nm,t }  \right)
    \label{eq:mechanistic_model}
\end{equation}

defined by:

\begin{itemize}[align=parleft,leftmargin=2cm,labelsep=.5cm]
    \item[$ P_{ n,t }$]  the probability of the person $n$ to receive an infectious dose. This shall not be understood as ``infection probability'', because~this strongly depends on the immune response by the affected person. 
    \item[$ \theta$]  the calibration factor for the specific disease.
    \item[$ \text{SR}_{m,t}$] the shedding rate, the~amount of virus the person $m$ spreads during the time step $t$.
    \item[$ i_{nm,t}$]  the intensity of the contact between $n$ and $m$, which corresponds to their distance.
    \item[$ t_{ nm,t }$]  the time the person $n$ interacts with person m during the time step $t$.
\end{itemize}

In our approach, we define the shedding rate $SR$ as a normalized bell-shaped function (Eq. \ref{eq:SR}) with $z \in (x, y)$ for both longitudinal and lateral dimensions, respectively (see \cite{schultz_evaluation_2020}). 

\begin{equation}    
    \label{eq:SR}
    \text{SR}_{xy} = \prod_{z \in (x, y)} \left( 1+ \frac{|z - c_z|}{a_z} ^{2 b_z} \right) ^{-1}
\end{equation}

\begin{figure}[htb!]
    \centering
    \includegraphics[width=.95\textwidth]{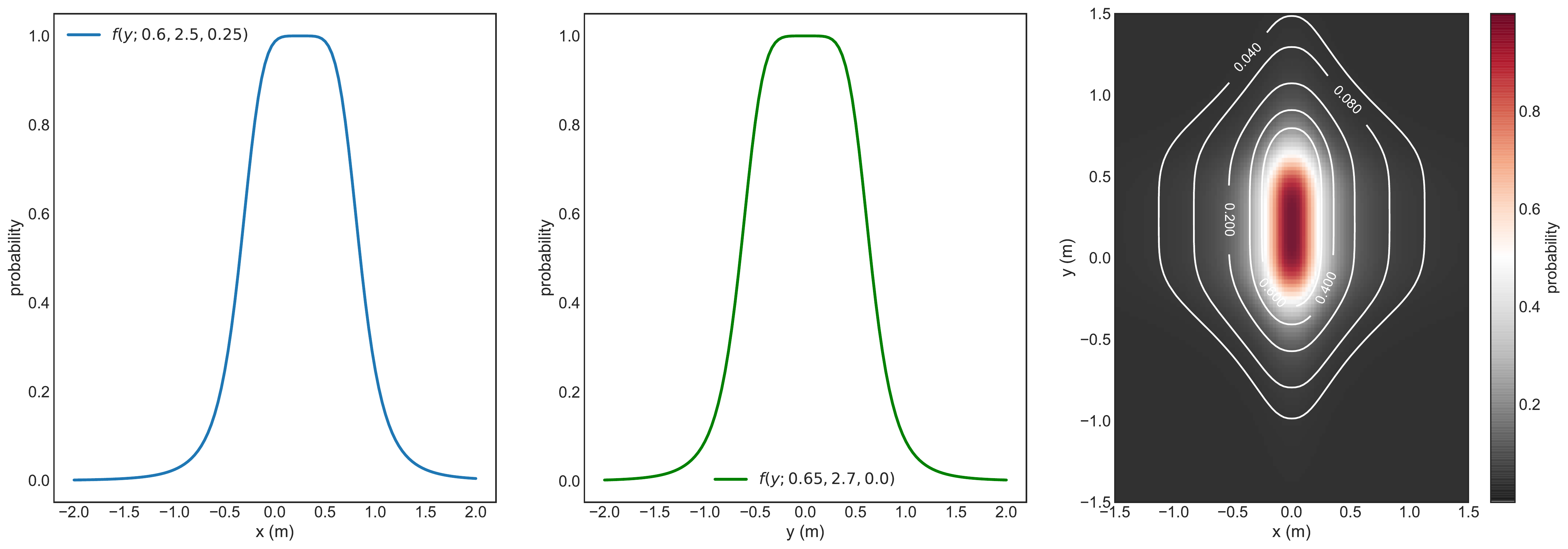}
    \caption{Transmission probability for longitudinal (x) and lateral (y) components and as two-dimensional probability field (right).}
    \label{fig:shedding_rate}
\end{figure}

The parameters are $a$ (scaling factor), $b$ (slope of leading and falling edge), and $c$ (offset) to determine the shape of the curve. The parameters have been initially set to $a_{x} = 0.6$, $b_{x} = 2.5$, $c_{x} = 0.25$, $a_{y} = 0.65$, $b_{y} = 2.7$, and $c_{y} = 0$. This generates a slightly smaller footprint in y-direction (lateral to moving direction) than in x-direction (longitudinal to moving direction). Additionally, the spread in x-direction is higher in front of the index case than behind it (see Fig. \ref{fig:shedding_rate}). Consequently, the moving direction is changed by 90 degrees with a heading to the aircraft window, when the passenger arrives his seat row.

Finally, the individual probability for virus transmission $P_n$ is corresponds to $\Theta$, the specific intensity (dose) per time step (Eq.~\ref{eq:pn_theta}). We set $\Theta$ to $\frac{1}{20}$, which means a passenger reaches a probability of $P_n = 1$ after standing 20~s in closest distance in front of an infected passenger (SR$_{xy}=1$). The parameter $\alpha \in \{1,2\}$ is 1 and changed to 2 when the passenger stores the luggage or enters the seat row. This doubled shedding rate reflects the higher physical activities within a short distance to surrounding~passengers.

\begin{equation}    
    P_n = \Theta \, \, \text{SR}_{xy} \, \, \alpha
    \label{eq:pn_theta}
\end{equation}

\subsection{Implementation and Analysis of Standard Boarding Approaches}
\label{sec:standard_boarding}

We introduce a baseline setup to depict the results for the evaluation of transmission risks, considering a seat load factor of 85\%, a conformance rate of 85\%, and an inter-arrival time of 3.7~s (exponential distributed) \cite{_Schultz2018c}. Tab. \ref{tab:infections} shows the comprehensive evaluation of transmissions around one infected passenger, which is randomly seated in the aircraft cabin. Two different scenarios are evaluated against the reference implementation (R) of the boarding strategies: (A) applying a minimum physical distance between two passengers of 1.6~m, and (B) additionally to the physical distance, the amount of hand luggage items is reduced by 50\%. Furthermore, the use of two aircraft doors in the front and at the rear is evaluated (A2 and B2) using the transmission risk and boarding time as indicators. In particular, the back-to-front strategy (2 blocks: front block with rows 1-15 , rear block with rows 16-29) exhibits lower values for the transmission probability than the optimized block strategy (using 6 blocks of aggregated seat rows) (see \cite{_Schultz2018c}). When passengers board (block-wise) from the back to the front, the chance to pass an infected person is reduced to a minimum, which is confirmed by the reduced transmission probability exhibited in Tab.~\ref{tab:infections}. This effect is also a root cause of the low transmission risks of the outside-in, reverse pyramid, and individual boarding strategy. 

\begin{table}[H]
    \caption{Evaluation of transmissions risk assuming one SARS-CoV2 passenger in the cabin. The simulated scenarios are: (R) reference implementation \cite{_Schultz2018c}, (A) 1.6~m minimum physical distance between two passengers, (B) additional reduction of hand luggage by 50\%, (A2) and (B2) use of two door configuration \cite{schultz_evaluation_2020}.}
    \label{tab:infections}
    \centering
    \begin{tabular}{l rrrrr c rrrrr}
    \toprule
     & \multicolumn{5}{c}{transmission risk} & & \multicolumn{5}{c}{boarding time (\%)}\\
     boarding strategy& R & A & B & A2 & B2 & &R & A & B & A2 & B2\\
    \hline
        random                      &   5.9 &   1.6 &   1.1 &   1.4 &   1.0 && 100  &    198 &   154 &133&103\\ 
    \midrule
        back-to-front (2 blocks)    & 	5.6 & 	1.4 & 	1.0 &   1.2 &   0.8 &&  96  &    220 &   169 &153&116\\ 
        optimized block (6 blocks)  & 	6.5 & 	2.3 &   1.5 &   1.5 &   1.0 &&  95  &    279 &   210 &166&125\\ 
    \midrule
        outside-in                  & 	3.5 & 	0.4 & 	0.2 & 	0.3 &   0.1 &&  80  &    161 & 	116 &107&77\\ 
        reverse pyramid             & 	3.0 & 	0.2 & 	0.1 &   0.2 &   0.1 &&  75  &    185 & 	128 &119&82\\ 
        individual                  & 	2.0 & 	0.2 & 	0.1 & 	0.2 &   0.1 &&  66  &    114 & 	104 &103&74\\ 
    \midrule
        deboarding                  & 	10.0 & 	9.7 & 	7.8 & 	7.6 &   6.0 &&  55  &    97  &   68  &52&36\\ 
    \bottomrule
    \end{tabular}
\end{table}

The use of two aircraft doors for boarding will provide an appropriate solution for a reduced transmission risk inside and outside the cabin, if near apron stands could be used and passengers could walk from the terminal to the aircraft. This kind of \textit{walk boarding} also prevents passengers from standing in the badly ventilated jetway during the boarding. Deboarding is difficult to control by specific procedures given that passengers demonstrated little discipline and high eagerness to leave the aircraft. More attention should be paid to this process and consideration should also be given to procedural or technical solutions to provide passengers better guidance and control.


\section{Optimized Boarding of Passenger Groups}
\label{sec:optimization}
In contrast to the prior analysed standard boarding procedures, we will provide a new model and a new optimization strategy which incorporates passenger groups and considers the requirements of physical distances in the aircraft cabin. In the following, we generally describe the mathematical problem and formulate the optimization model.

\subsection{Problem Description and Formulation}
We develop a new mathematical model to determine an optimal strategy for assigning seats in cabin under the objective to minimize the virus transmission risk. The idea to create an appropriate seat allocation for a pandemic situation includes three assumptions. The first one is that an airline could assign just a percentage of the available seats (e.g. 50\%) to reduce the virus transmissions in cabin and this strategy will be the primary solution to face with the pandemic situation. The next assumption is about minimizing passenger contacts or maximizing the distances between passengers in the cabin and guaranteeing at the same time that the confined space inside the aircraft is used efficiently. 

\begin{figure}[htb!]
    \centering
    \includegraphics[width=.7\textwidth]{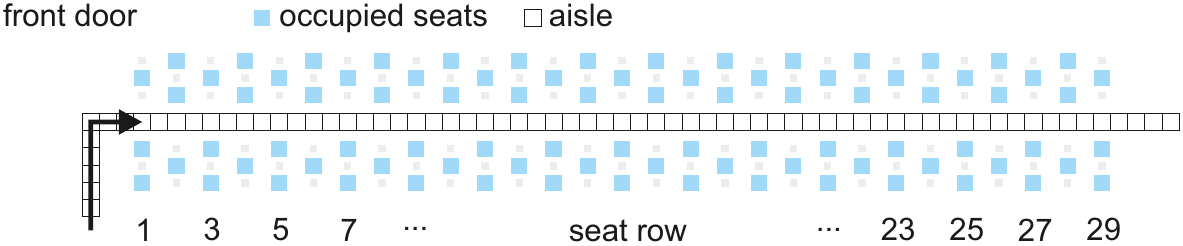}
    \caption{Fifty percent of the seats will be allocated to passengers during the pandemic situation according to a pattern with maximum physical separation.}
    \label{fig:50percentPax}
\end{figure}

Looking at Fig.~\ref{fig:50percentPax}, passengers have maximized distances from each other respecting the limitation that only 50\% of the seats can be occupied. Each airline company could determine the seat load factor for each flight individually also considering risk assessments or economic reasons. Indeed, many airlines are currently operating the generally accepted strategy of the empty middle seat.

Although complex boarding strategies, such as outside-in, reverse pyramid and individual lead to better boarding times, there will be an issue. The boarding process is driven by the willingness of passengers to follow the proposed strategy. We will assume a group of four members (e.g. a family) to be seated. If one of these boarding strategies are applied, they will have just two options. The first one is seating near each other, therefore they have to split during the boarding (see Fig.~\ref{fig:boarding_strategies}). The next option is remaining as one group in the boarding sequence and as a result they have to seat in different rows. Both options are inconvenient for group members (families). Here we propose to look at the group members as a community, since they were already in close contact before boarding. The strategy that is used in Fig.~\ref{fig:50percentPax} depicts a general solution, but it could be improved considering groups. Without loss of generality, we could suppose that the transmission rate for the members of each group is zero, which will result in a better use of space and create a new pattern. 

\begin{figure}[htb!]
    \centering
    \includegraphics[width=.9\textwidth]{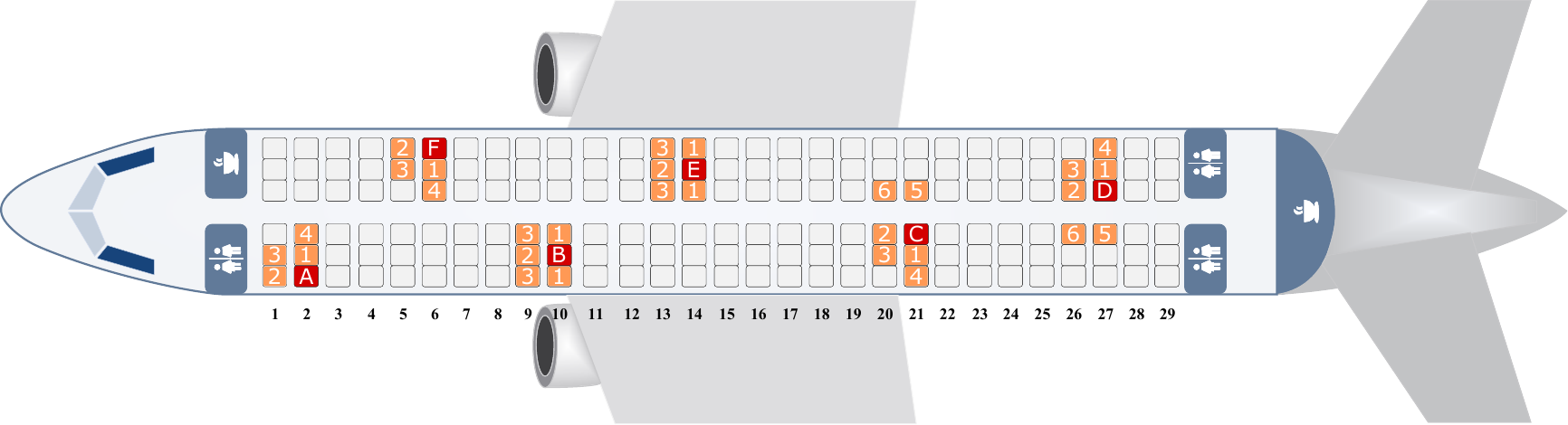}
    \caption{Different types of interactions generated around the infected passengers (coded red).}
    \label{fig:interactions}
\end{figure}

The introduced concept of a shedding rate of infected passengers will be used here as well. If an infected passenger was assigned to different columns, the several shedding rates must be counted based on the location of the adjacent locations. Taking Fig.~\ref{fig:interactions} as an example, when a passenger seated in row $i=21$ and column C (aisle), we compute the shedding rate for the passenger from other groups that seat in the same row ($i=21$ at column A (window), B (middle), and D (aisle)) and previous row $i-1=20$ (column B (middle), C (aisle), and D (aisle)). 

\subsection{Optimization Model}
Based on the assumptions of the problem description, we list the sets, parameters, and decision variables for achievement of the purposes of the research.

\begin{table}[H]
    \centering
    
    \begin{tabular}{lll}
        \hline
        \textbf{Notation} &  \textbf{Definition} \\
        \emph{Sets and Indexes} & \\
        $i$ & Index set of row $i\in \{1,2, \dots, \mathcal{I}\}$\\
        $j$ & Index set of column $j\in \{1,2, \dots, \mathcal{J}\}$\\
        $k$ & Index set of group $k\in \{1,2, \dots,\mathcal{K}\}$\\
        $r$ & Index set of interaction type $r\in \{1,2, \dots, \mathcal{R}\}$\\
        ~\\
        
        \emph{Parameters} & \\
        $T_{k}$ & Number of members in the group $k$ \\
        $SR_{r}$ & The related shedding rate for interaction $r$\\
        ~\\
        
        \emph{Decision Variables} & \\
        $x_{ijk}$ & Binary variable, equals one if a passenger from group $k$ is seated in a seat\\
        & in row $i$ and column $j$; equals zero otherwise\\
        $d_{ijk}$ & The summation of shedding rates that the passengers of other groups can cause\\
        & for a passenger from group $k$ who is seated in a seat in row $i$ and column $j$\\
        
        \hline
    \end{tabular}
    \label{tab:param}
\end{table}

The proposed a mixed-integer linear programming model for the problem is introduced as follows.

\begin{align}
 \hspace{10pt}  min\hspace{4pt} \sum_{i=1}^{\mathcal{I}}\sum_{j=1}^{\mathcal{J}}\sum_{k=1}^{\mathcal{K}}d_{ijk}\label{eq:objFunc}
\end{align}
\begingroup
    \allowdisplaybreaks
    \begin{align}
        & \sum_{k=1}^{\mathcal{K}}x_{ijk} \le 1  & \forall i,j \label{eq:eqa5}\\
        & \sum_{i=1}^{\mathcal{I}}\sum_{j=1}^{\mathcal{J}} x_{ijk} = T_{k}  & \forall k \label{eq:eqa6}\\
        & 6(x_{ijk}-1) + \sum_{k'=1, k\neq k'}^{\mathcal{K}} \{ SR_{1}x_{i(j+1)k'} +  SR_{4}x_{i(j+2)k'} \} \le d_{ijk}  & \forall i=1,j=1,k \label{eq:eqa7}\\
        & 6(x_{ijk}-1) + \sum_{k'=1, k\neq k'}^{\mathcal{K}} SR_{1} \{ x_{i(j-1)k'} + x_{i(j+1)k'} \} \le d_{ijk}  & \forall i=1,j=2,5, \, \, k \label{eq:eqa8}\\
        & 6(x_{ijk}-1) + \sum_{k'=1, k\neq k'}^{\mathcal{K}} \{ SR_{4}x_{i(j-2)k'} + SR_{1}x_{i(j-1)k'} + SR_{5}x_{i(j+1)k'} \} \le d_{ijk}  & \forall i=1,j=3,k \label{eq:eqa9}\\
        & 6(x_{ijk}-1) + \sum_{k'=1, k\neq k'}^{\mathcal{K}} \{ SR_{5}x_{i(j-1)k'} + SR_{1}x_{i(j+1)k'} + SR_{4}x_{i(j+2)k'} \} \le d_{ijk}  & \forall i=1,j=4,k \label{eq:eqa10}\\
        & 6(x_{ijk}-1) + \sum_{k'=1, k\neq k'}^{\mathcal{K}} \{ SR_{4}x_{i(j-2)k'} + SR_{1}x_{i(j-1)k'}\}  \le d_{ijk}  & \forall i=1,j=6,k \label{eq:eqa11}\\
        & 6(x_{ijk}-1) + \sum_{k'=1, k\neq k'}^{\mathcal{K}} \{ SR_{2}x_{(i-1)jk'} +  SR_{3}x_{(i-1)(j+1)k'} + SR_{1}x_{i(j+1)k'} \nonumber \\
        & \quad\quad\quad\quad\quad\quad\quad\quad\quad + SR_{4}x_{i(j+2)k'} \}  \le d_{ijk} & \forall i\ge 2,j=1,k \label{eq:eqa12}\\
        & 6(x_{ijk}-1) + \sum_{k'=1, k\neq k'}^{\mathcal{K}} \{ SR_{3}x_{(i-1)(j-1)k'} + SR_{2}x_{(i-1)jk'}+ SR_{3}x_{(i-1)(j+1)k'}
        \nonumber \\
        & \quad\quad\quad\quad\quad\quad\quad\quad\quad+ SR_{1}x_{i(j-1)k'}+ SR_{1}x_{i(j+1)k'} \} \le d_{ijk}  &\forall i\ge 2,j=2,5, \, \, k \label{eq:eqa13}\\
        & 6(x_{ijk}-1) + \sum_{k'=1, k\neq k'}^{\mathcal{K}} \{ SR_{3}x_{(i-1)(j-1)k'} + SR_{2}x_{(i-1)jk'}+ SR_{6}x_{(i-1)(j+1)k'}
        \nonumber \\
        & \quad\quad\quad\quad\quad\quad\quad\quad\quad + SR_{4}x_{i(j-2)k'}+ SR_{1}x_{i(j-1)k'}+ SR_{5}x_{i(j+1)k'}\} \le d_{ijk}  & \forall i\ge 2,j=3,k \label{eq:eqa14}\\
        & 6(x_{ijk}-1) + \sum_{k'=1, k\neq k'}^{\mathcal{K}} \{SR_{6}x_{(i-1)(j-1)k'} + SR_{2}x_{(i-1)jk'}+ SR_{3}x_{(i-1)(j+1)k'}
        \nonumber \\
        & \quad\quad\quad\quad\quad\quad\quad\quad\quad + SR_{5}x_{i(j-1)k'}+ SR_{1}x_{i(j+1)k'}+ SR_{4}x_{i(j+2)k'} \}  \le d_{ijk}  & \forall i\ge 2,j=4,k \label{eq:eqa15}\\
        & 6(x_{ijk}-1) + \sum_{k'=1, k\neq k'}^{\mathcal{K}} \{SR_{3}x_{(i-1)(j-1)k'} + SR_{2}x_{(i-1)jk'}+SR_{4}x_{i(j-2)k'}
        \nonumber \\
        & \quad\quad\quad\quad\quad\quad\quad\quad\quad + SR_{1}x_{i(j-1)k'}\} \le d_{ijk}  & \forall i\ge 2,j=6,k \label{eq:eqa16}\\
        & x_{ijk}\in \{0,1\}, \quad d_{ijk}\ge 0  & \forall i,j,k \label{eq:eqa17}
    \end{align}
\endgroup

The summation of shedding rates of all passengers, as objective function, is minimized in equation~(\ref{eq:objFunc}). Constraints (\ref{eq:eqa5}) guarantee that each seat would be assigned to not more than one passenger. The number of group members are indicated by constraints (\ref{eq:eqa6}). Constraints (\ref{eq:eqa7})-(\ref{eq:eqa11}) correspond to the shedding rates of passengers that are seated in the first row~(i=1) in cabin. For instance, if a passenger was seated in a seat in column C~(j=3), then the shedding rate for that passenger can be calculated based on the other passengers of different groups that were seated in columns A~(j=1), B~(j=2), and D~(j=4) on constraints (\ref{eq:eqa9}). Also, the shedding rates of passengers that are seated in other rows in cabin are computed by constraints (\ref{eq:eqa12})-(\ref{eq:eqa16}). Here, we must consider the shedding rates not only for passenger in the same row, but also for the previous row. For instance, if a seat in the second row which located in column C~(j=3), was assigned to a passenger, then the shedding rates of the other passengers from different groups that were seated in column A~(j=1), B~(j=2), and D~(j=4) of that row and columns B~(j=2), C~(j=3), and D~(j=4) of the first row are calculated based on constraints (\ref{eq:eqa14}) as well. On the left hand side of the constraints (\ref{eq:eqa7})-(\ref{eq:eqa16}), the first term takes a value of zero if the seat (i, j) is assigned to group $k$ and $-6$ otherwise. As a result, when there is not a passenger in a seat (i,j), the shedding rate takes zero. Finally, the requirements for decision variables are denoted by constraints (\ref{eq:eqa17}).

\section{Application of the Model and Evaluation of the Results}
\label{sec:application_optimization}
We solved the mathematical model for a small size problem~($i,j,k=6$). When we increased the size of problem to the medium size~(e.g. $i,j,k=10$), the time run increased significantly and the optimization software (GAMS with CPLEX solver) could not find an optimal solution in a reasonable time~(10 hours). Therefore, we suggested to design a Genetic Algorithm (GA) for the real sized problem. The problems run on a computer with the specifications the AMD Ryzen 7, 3700U, 2.30GHz CPU, 16 GB RAM, and Matlab 2013 software is used for running the GA. The developed model is applied to derive an optimal seat allocation using a genetic algorithm. We choose five use cases for the optimization and implemented the optimal seat allocation in the passenger boarding simulation to provide an appropriate boarding strategy.

\subsection{Solution Procedure and Result}
\label{sec:model's result}
The GA has many applications in the optimization problems. In light of the NP-hard class of the seat layout and boarding problem, several methods were conducted to present the optimal/near optimal solutions \cite{van_den_briel_america_2005, majid_soolaki_2012}. Therefore, we designed an GA to solve the problem. The proposed chromosome structure is represented as follows:

\[
  C=
  \left[ {\begin{array}{cccccc}
   y_{1,1} & y_{1,2} & y_{1,3} & y_{1,4} & y_{1,5} & y_{1,6}\\
   y_{2,1} & y_{2,2} & y_{2,3} & y_{2,4} & y_{2,5} & y_{2,6}\\
   ... & ... & ...& ... & ... & ... \\
   y_{29,1} & y_{29,2} & y_{29,3} & y_{29,4} & y_{29,5} & y_{29,6}\\  
  \end{array} } \right]\\
  \quad , y_{i,j}=k, \quad \text{if} \quad x_{i,j,k}=1, \text{otherwise zero}.\\
\] 

The value of each array of the matrix such as $y_{i,j}$ is the group's number of related decision variable $x_{i,j,k}$, which is $k$ if the seat (row $i$, column $j$) is assigned to a passenger group $k$, otherwise it takes zero. The population of chromosomes for the first generation are created based on the structure above. Also, we evaluate each chromosome with it's fitness function value which is determined by the value of the original objective function. After that, we implement several operators as follows to create the next generations from the current generation: selection, crossover, mutation, migration, and elitism operator.

\begin{figure}[htb!]
    \centering
    \includegraphics[width=.35\textwidth]{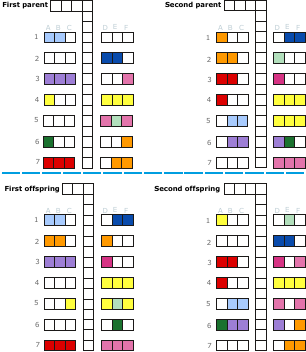} 
    \hspace{50pt}
    \includegraphics[width=.35\textwidth]{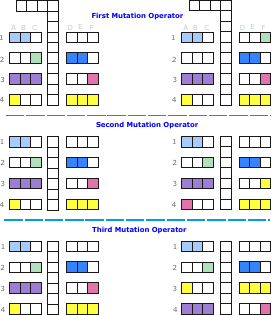}
    \caption{Crossover operator (left) and mutation operators (right).}
    \label{fig:CMoperators}
\end{figure}

The selection operator (roulette wheel) ensures that each chromosome with a lower fitness function value is more likely to be selected. New offsprings are created by a recombination of parental genes. Therefore, the group's numbers are divided into two separate sets. The first offspring receives their genes (the value of arrays in Matrix $C$) of the first set from the first parent and the second part from the second parent and vice versa for the second offspring (see Fig.~\ref{fig:CMoperators} (left)). Here, we explain with colors to clarify the implication of operation. For example, the first offspring receives their genes which are colored with light green, light blue, purple, and red from the first parent and receives the genes which are colored with dark green, navy, pink, orange, and yellow from the second parent. If there is overlap between a gene's location of the first and second parent, then we use a random strategy to select another array in matrix and value it~(e.g. for the first offspring, instead of $y_{5,5}=31$, we  randomly set $y_{5,3}=31$ because the array~(5,5) was colored with light green before or $y_{5,5}=6$).

Therefore, we always create feasible solutions. The mutation operator is used to maintain the diversity of solutions. Therefore, we designed several operators to change a number of genes in a chromosome to create a new chromosome for the next generation (see Fig.~\ref{fig:CMoperators} (right)). In the first mutation operator, we change the seat location of a passenger~($y_{2,3}=6$), as a gene in chromosome, to an unoccupied seat~($y_{1,6}=0$). In the second operator, we change the locations of two occupied seats~($y_{4,1}=31$ and $y_{3,6}=24$ ). Therefore, after implementation of the operator, we have: $y_{4,1}=24$ and $y_{3,6}=31$. Finally, the arrays of two random rows (the third and fourth rows) are changed in the last mutation operator. A low percentage of each generation is randomly transferred to the next generation (migration operator). The elitism operator selects the best chromosomes in terms of fitness function value, and transfer them form the current generation to the next generation. The following parameters were used for executing the code: initial population = 1000, number of generations = 1000, crossover rate = 0.55, mutation rate = 0.35, elitism = 0.075, and migration rate = 0.025.

\begin{figure}[htb!]
    \centering
    \includegraphics[width=.9\textwidth]{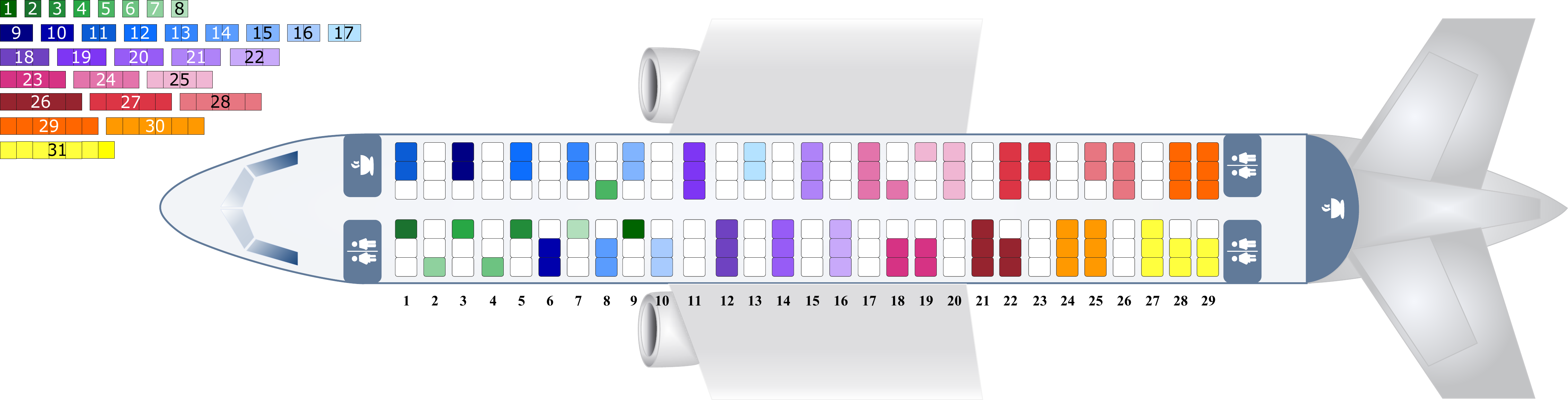}
    \caption{Best layout to seat 31 groups (87 passengers, 50\% seat load) solved by GA approach.}
    \label{fig:GA_scenrario_3}
\end{figure}

We consider 8 groups with one member (i.e. G1 to G8 which coded with green color), 9 groups of two members (i.e. G9 to G17 which coded with blue color), 5 groups of three members (i.e. G18 to G22 which coded with purple color), 3 groups of four members (i.e. G23 to G25 which coded with ping color), 3 groups of five members (i.e. G26 to G28 which coded with red color), 2 groups of six members (i.e. G29 to G30 which coded with orange color), and finally a group of seven members (i.e. G31 which coded with yellow color). Fig.~\ref{fig:GA_scenrario_3} depicts an optimized solution for the seating layout based on the designed GA generated for the 31 groups (87 passengers). 

\begin{figure}[htb!]
    \centering
    \includegraphics[width=.9\textwidth]{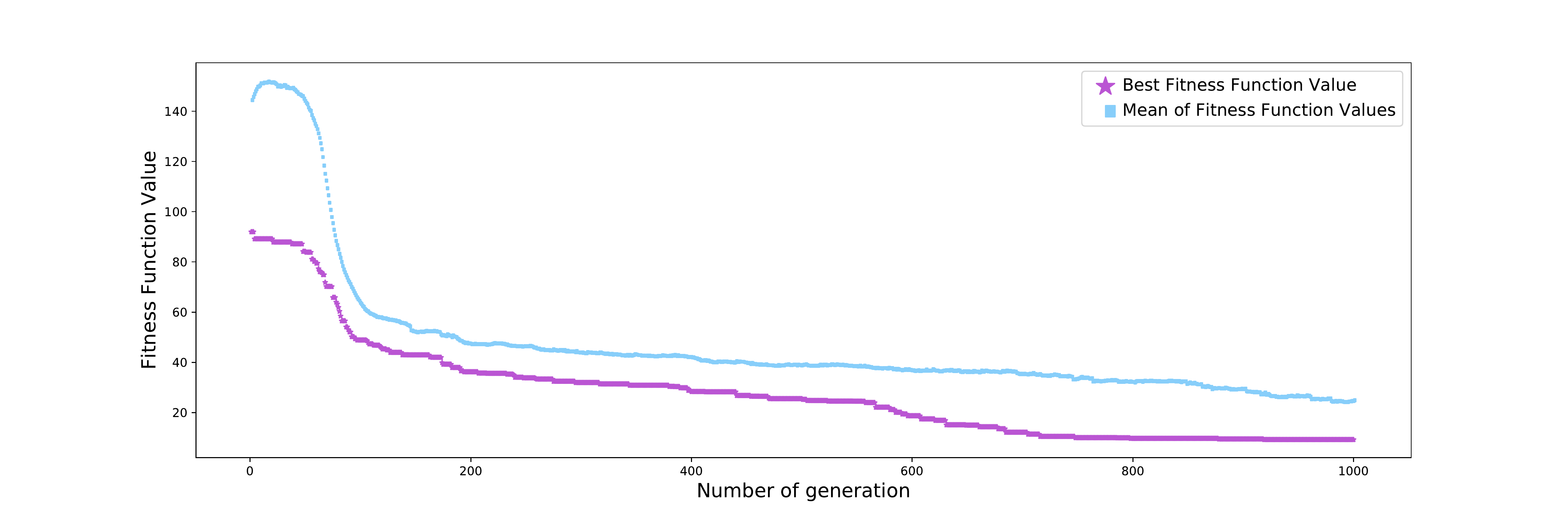}
    \caption{Progress of GA fitness function.}
    \label{fig:evo_diagram}
\end{figure}

The run time for GA is 1805 s, the value of the objective function for the best solution is 9.1916. The evolutionary diagram concerning the GA is shown in Fig.~\ref{fig:evo_diagram}. The fitness function of the elite and the mean of each generation demonstrate the increasing quality of generated solutions (decreasing fitness function).

To understand the impact of our group approach, we introduce five scenarios and their related solutions based on the assumptions below and compare them by the values of the objective function and the number of passengers considered.

\begin{itemize}
    \item Scenario 1: Aircraft seats are assigned randomly to passengers with a maximum distance and a seat load of 50\% (87 passengers).
    \item Scenario 2: Similar with scenario 1, while the group members are seated close to each other in the same area.
    \item Scenario 3: Optimized solution from mathematical modelling and GA application (indicated in Fig.~\ref{fig:GA_scenrario_3}).
    \item Scenario 4: Optimized solution considering increased seat load of 66\% (115 passengers).
    \item Scenario 5: Optimized solution considering for maximum number of passengers (174).   
\end{itemize}    

\begin{figure}[htb!]
    \centering
    \includegraphics[width=.9\textwidth]{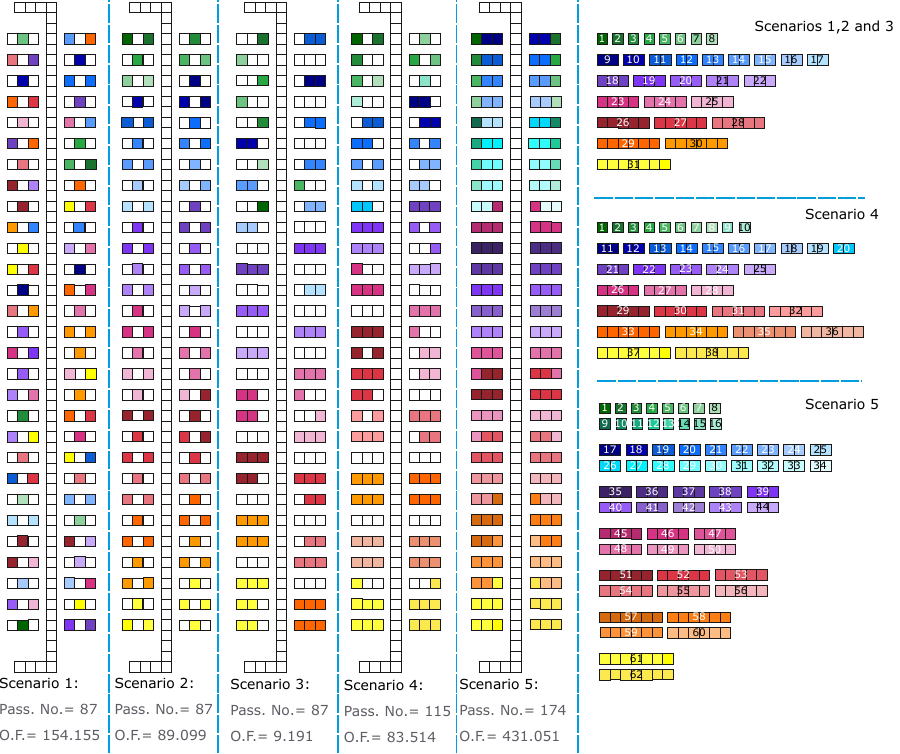}
    \caption{Five different solutions for optimized seat allocation in the aircraft cabin considering 87, 115, and 174 passengers assigned to different groups.}
    \label{fig:five_cases}
\end{figure}

The corresponding solutions for the scenarios are illustrated in Fig.~\ref{fig:five_cases}. In the scenarios 1 to 3, the number of passengers is fixed to 87 passengers. The values of the objective function (O.F.) of these three scenarios exhibit that our approach (scenario 3) for an optimized seat layout results in a significantly reduction of the transmission risk: a reduction of 94\% compared to scenario 1 (seats are assigned randomly to passengers with a maximum distance), and a reduction of 90\%  compared to scenario 2 (seats are assigned randomly, group members in the same zone). In addition, the optimization method uses the available space in the best way, so airlines could benefit from our approach. For example, although we were increasing the number of passengers by 33\% and created the scenario 4, the objective function of that scenario is still lower than the second case. Finally, the consideration of a seat load of 100\% indicates an upper boundary (scenario~5) for the objective function.

\begin{figure}[htb!]
    \centering
    \includegraphics[width=.9\textwidth]{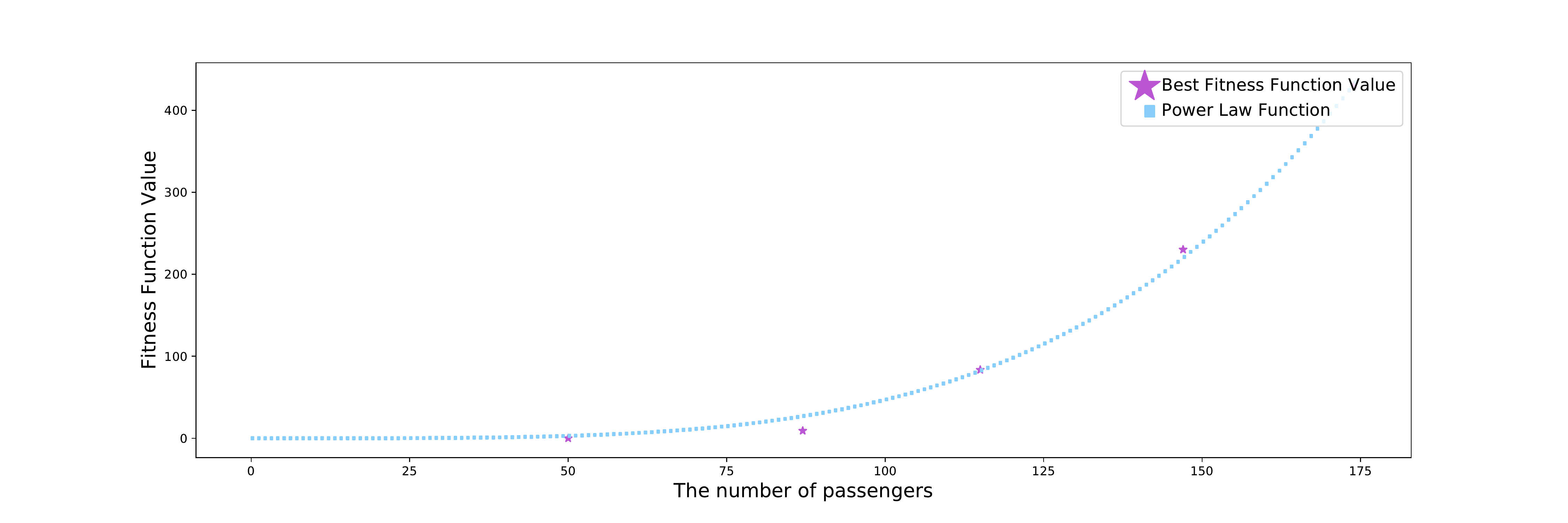}
    \caption{Progressive increase of transmission risk over the number of seated passengers.}
    \label{fig:progressive}
\end{figure}

If the seat load increases over 50\% (87 passengers) the values of the objective function (transmission risk) progressively increases as shown in Fig.\ref{fig:progressive}. Assuming an average seat load of 85\% (147 passengers) airlines could significantly reduce the transmission risk by two third by implementing our group approach and a reduced seat load of 66\% (115 passengers). To show the general behaviour of the objective function, we use a power law function $y=ax^b$ with $a=4.705083 \times 10^{-7}$ and $b=4.001411$.

\subsection{Boarding times and evaluation}
The implementation of the mixed-integer linear programming model and the genetic algorithm result in an optimized layout for the passengers to be seated in the aircraft cabin. This layout will be used as input for the passenger boarding model, which was extended by a transmission module to evaluate transmission risk during aircraft boarding, to derive an optimum sequence to board the passengers. In our contribution, we will not provide an optimization of the deboarding process. 

\begin{figure}[htb!]
    \centering
    \includegraphics[width=.9\textwidth]{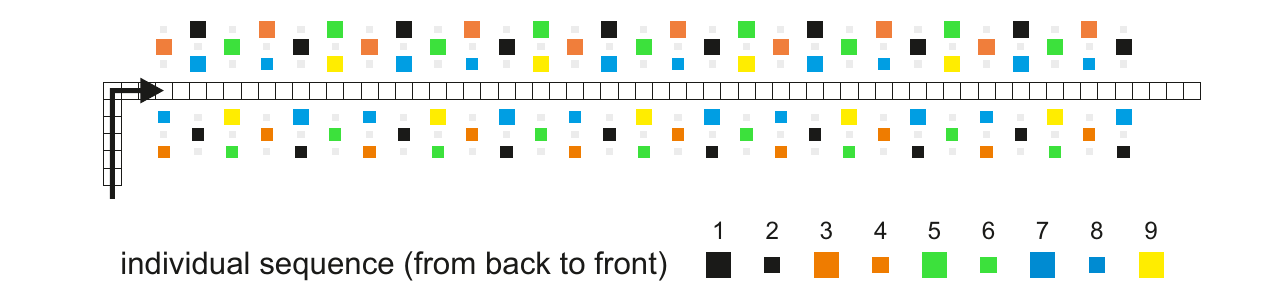}
    \caption{Optimized individual boarding sequence considering a physical distance of 1.6~m between passengers (scenario 1).}
    \label{fig:individual-boarding_opt_physDist}
\end{figure}

Analyses in the context of appropriate boarding sequence accompanied by the introduction of infrastructural changes showed that an optimized sequence comprises a mix of boarding per seat (from window to aisle) and per seat row (from the rear to the front) \cite{_schultz2017a}. First and foremost, per-seat boarding (window seats first) is the most important rule to ensure seating without additional interaction in the seat rows. Starting with an outer seat in the last row, the number of group members and the necessary physical distance between passengers (1.6~m) defines the subsequently following seat row, which could be used in parallel (e.g. 6 passengers with seat row 29 will block  the aisle until seat row 27 (waiting), the physical distance requires to block row 26 and 25, the next group must have seats in front of row 25). This process of seat and row selection is repeated until the front of the aircraft is reached and is repeated until all passengers are seated. We further assume that the passengers in each group will organize themselves appropriately to minimize local interactions. In Fig.~\ref{fig:individual-boarding_opt_physDist}, the result of this sequencing algorithm is exemplary illustrated.

\begin{figure}[htb!]
    \centering
    \includegraphics[width=.9\textwidth]{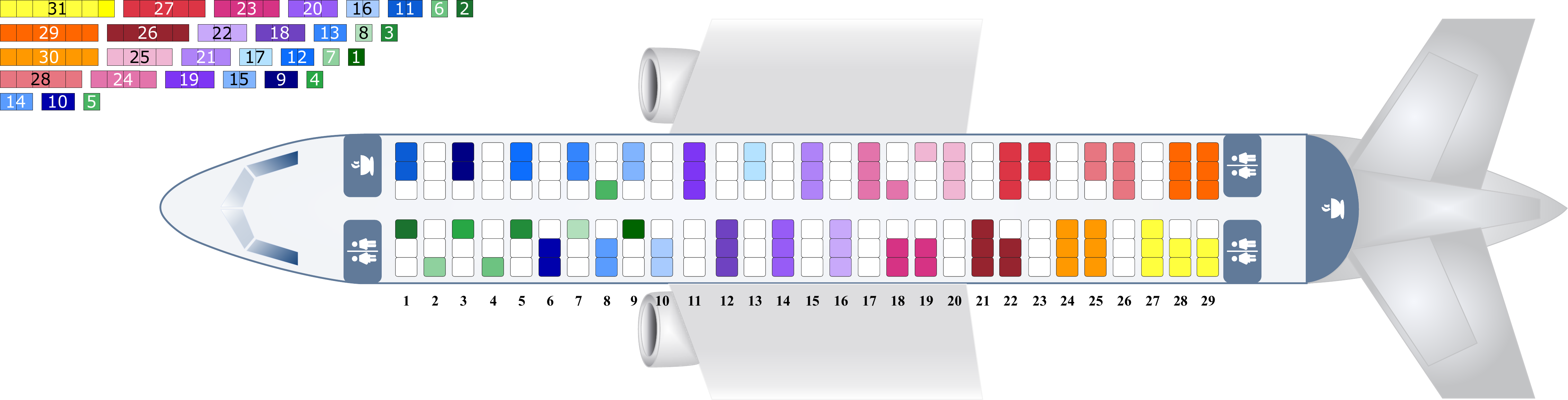}
    \caption{Optimized individual boarding sequence considering a physical distance of 1.6~m between passengers (scenario 3).}
    \label{fig:scenario3_sequence}
\end{figure}

If the sequencing algorithm is applied to the optimized seat layout from scenario~3, the passenger groups are boarded in five segments. Inside each group, the distance between passengers is not restricted but between groups it is constrained by 1.6~m (last member of the first group and the first member of the following group). The first segment starts with group no.~31 and the last segment with group no.~14 (see Fig.~\ref{fig:scenario3_sequence}). As an example, the passengers inside group no.~31 (yellow) are organized by the following sequence of seats, which results in a minimum of individual seat and row interactions: 29A, 29B, 28A, 28B, 27A, 27B, and 27C. Considering distances between groups, the best candidate will be group no.~27 (red) with the seats 23F, 23E, 22F, 22E, and 22D. This sequence allows both groups to start the seating process in parallel, without waiting time due to a too small distance between the seat rows.

In the first three scenarios 87 passengers are boarded with different strategies (see Fig.~\ref{fig:five_cases}): individual passengers in a regular pattern (scenario 1), groups in a regular pattern (scenario 2), and groups in an optimized seat layout (scenario 3). Scenario 1 is used as reference case to evaluate the performance (boarding time) and the transmission risk of Scenario 2 and 3 (Tab.~\ref{tab:boarding_sim_results_a}). Therefore, the passenger sequence is established for both random and individual boarding strategy (optimized, Fig.~\ref{fig:individual-boarding_opt_physDist}). The boarding time for the random strategy is set to 100\%. As shown in Tab.~\ref{tab:boarding_sim_results_a}, the implementation of the individual strategy will reduce the boarding time to 45.2\% at a minimum of transmission risk. The consideration of groups (scenario 2 and 3) using the random strategy already reduces the boarding time by about a third at a comparable level of transmission risk. If the optimized seat allocation is used together with the individual (group) boarding the boarding time could be further reduced to 41.1\% at a low transmission risk of 0.09 new infected passengers at average (85\% reduction). Tab.~\ref{tab:boarding_sim_results_b} emphasizes the portability of the results achieved by the evaluation of scenarios 4 and 5. A corresponding baseline was calculated for each scenario (random boarding of individual passengers). The boarding of groups reduces the boarding time and the transmission risk, and the optimization of the boarding sequence additionally leads to a significant reduction of the transmission risk of about 65\%. 

\begin{table}[!htb]
    \caption{Evaluation of average aircraft boarding times and transmission risk for one randomly seated infectious passenger.}
    \begin{subtable}{.5\linewidth}
      \centering
        \caption{Scenario 1,2, 3 with 87 passengers.}
        \begin{tabular}{l l rr}
            \toprule
            Sce- & Strategy & Boarding& Transmis-\\
            nario& & Time (\%) & sion risk\\
            \toprule
             1  & random            & 100.0 & 0.58\\ 
                    & individual        & 45.2  & 0.00\\
             2  & groups, random    & 68.0  & 0.62\\
                    & groups, individual      & 51.9  & 0.20\\
             3  & groups, random    & 69.0  & 0.57\\
                    & groups, individual      & 41.1  & 0.09\\ 
            \bottomrule
        \end{tabular}
        \label{tab:boarding_sim_results_a}
    \end{subtable}%
    \ 
    \begin{subtable}{.5\linewidth}
      \centering
        \caption{Scenario 4 (115 passengers) and 5 (174 passengers)}
        \begin{tabular}{ll rr}
            \toprule
            Sce- & Strategy & Boarding& Transmis-\\
            nario& & Time (\%) & sion risk\\
            \toprule        
                4  & random            & 100.0 & 1.11\\ 
                    & groups, random    & 60.5  & 0.94\\
                    & groups, individual      & 38.1  & 0.31\\
                5  & random            & 100.0 & 2.09\\
                    & groups, random    & 65.1  & 1.96\\
                    & groups, individual      & 34.4  & 0.66\\ 
        \bottomrule
        \end{tabular}
        \label{tab:boarding_sim_results_b}
    \end{subtable} 
    \label{tab:boarding_sim_results}
\end{table}															
	
Finally, our results show that optimized group boarding of 174 passenger (scenario 5) possesses a transmission risk of 0.66, which is close to the random strategy in scenario 1 (0.58). Furthermore, this particular scenario 5 performs about 20\% faster than the random strategy in scenario 1 (87 passengers) and reaches pre-pandemic boarding times.




\section{Discussion and outlook}
\label{sec:discussion}

Along the passenger journey, the processes in the aircraft cabin require sharing a confined environment with other passengers during boarding and flight. These processes have the risk of virus transmission between passengers and require appropriate seat configuration and risk mitigation strategies. A physical distance between passengers during boarding and staggered seat configurations are part of the risk mitigation strategy. However, the side effect from an operational point of view is a doubled boarding time compared to the situation before the coronavirus pandemic situation.

In our contribution, we consider passenger groups as an important factor for the operational efficiency. The main idea behind our approach is that members of one group are allowed to be close to each other, as they already are before boarding, while different groups should be as far apart as necessary. We provide a customer-oriented solution for both airlines and passengers, which enables a situative approach to establish appropriate seat allocation and aircraft entry sequences considering a minimum transmission risk between groups of passengers. Thus, we developed a new mathematical model, which provides an optimized seat allocation, while minimizing the sum of shedding rates that an infected passenger can cause. The developed model was used to evaluate the transmission risk of a seat allocation scheme and to solve this optimization problem with a genetic algorithm for three different scenarios of grouped passengers (87, 115, 174). The optimization of a standard scenario with a seat load of 50\% (87~passengers) shows that with the consideration of groups the value for the objective function was reduced from 154 to 9, which means a significant reduction of the transmission risk induced by the new seat allocation. 

Five seat and group configurations were used as input for the boarding simulation (stochastic cellular automata), which evaluates the transmission risk during the passenger movements in the cabin (walk the aisle, store luggage, take the seat). Therefore, the sequence of groups were optimized to keep the boarding time as low as possible. Our simulation results exhibit that the optimized seat allocation for groups (scenario 3) performs best for the boarding time (41.1\% in relation to random boarding with no groups) at a low level of transmission risk (0.09, while random boarding without groups leads to a risk of 0.58). We could also demonstrate that the effective consideration of passenger groups is a major impact factor for fast and safe passenger boarding (e.g., board more passengers at the same level of transmission risk). In the context of aircraft ground operations (turnaround), shorten boarding times could compensate the extended ground times caused by additional disinfection procedures in the aircraft cabin.

\vspace{6pt}

\conflictsofinterest{The authors declare no conflict of interest.}

\reftitle{References}
\externalbibliography{yes}

\bibliography{biblio,mst}

\end{document}